\documentclass[aps,prl,reprint,twocolumn,superscriptaddress]{revtex4-1}
\usepackage[caption=false]{subfig}
\usepackage{amsmath,amsfonts,amsthm,graphicx,mathtools,epstopdf,array,etoolbox,hyperref,makecell}
\usepackage[T1]{fontenc}

\usepackage{pdfpages}
\makeatletter 
\AtBeginDocument{\let\LS@rot\@undefined}
\makeatother
\usepackage{pgffor} 

\newcommand{\ket}[1]{\left| #1 \right>}
\newcommand{\bra}[1]{\left< #1 \right|}
\newcommand{\ketbra}[2]{\ket{#1} \! \bra{#2}}
\newcommand{\binh}{h_2} 

\newcommand{\pr}{\mathrm{Pr}} 
\newcommand{\prt}{\pr_\mathrm{target}} 
\newcommand{\norm}[1]{\left\lVert #1 \right\rVert}
\newcommand{\e}{\epsilon} 
\newcommand{\q}{q} 
\newcommand{\qt}{\q_t} 
\newcommand{\etat}{\eta_t} 
 
\newcommand{\str}[1]{\boldsymbol{#1}} 
\newcommand{\flip}[1]{\overline{#1}} 
\newcommand{\symmset}{\mathcal{S}}

\newcommand{\defvar}{\coloneqq} 

\newcommand{\scenoneway}{i}
\newcommand{\scenHardyopt}{ii}
\newcommand{\scenMY}{iii}
\newcommand{\scenCHSHbasic}{iv}
\newcommand{\scenCHSHq}{v}
\newcommand{\scenCHSHeta}{vi}

\newcommand{\appsymm}{C}

\newcommand{\appattack}{E}
\newcommand{\appfvdg}{F}

\newtheorem{theorem}{Theorem}

\newtheorem{corollary}{Corollary}

\newcolumntype{L}[1]{>{\raggedright\let\newline\\\arraybackslash\hspace{0pt}}m{#1}}
\newcolumntype{C}[1]{>{\centering\let\newline\\\arraybackslash\hspace{0pt}}m{#1}}
\newcolumntype{R}[1]{>{\raggedleft\let\newline\\\arraybackslash\hspace{0pt}}m{#1}}

\newcounter{tabrownumcounter}
\newcommand\tabrownum{\stepcounter{tabrownumcounter}\roman{tabrownumcounter}}

\begin{document}

\title{\textbf{Advantage distillation for device-independent quantum key distribution}}

\author{Ernest Y.-Z.\ Tan}
\affiliation{Institute for Theoretical Physics, ETH Z\"{u}rich, Switzerland}
\author{Charles C.-W.\ Lim}
\affiliation{Department of Electrical \& Computer  Engineering, National University of Singapore, Singapore}
\affiliation{Centre for Quantum Technologies, National University of Singapore, Singapore}
\author{Renato Renner}
\affiliation{Institute for Theoretical Physics, ETH Z\"{u}rich, Switzerland}

\begin{abstract}
Device-independent quantum key distribution (DIQKD) offers the prospect of distributing secret keys with only minimal security assumptions, by making use of a Bell violation. However, existing DIQKD security proofs have low noise tolerances, making a proof-of-principle demonstration currently infeasible. We investigate whether the noise tolerance can be improved by using advantage distillation, which refers to using two-way communication instead of the one-way error-correction currently used in DIQKD security proofs. We derive an efficiently verifiable condition to certify that advantage distillation is secure against collective attacks in a variety of DIQKD scenarios, and use this to show that it can indeed allow higher noise tolerances, which could help to pave the way towards an experimental implementation of DIQKD.
\end{abstract}

\maketitle

\emph{Introduction} --- 
In quantum key distribution, the goal is to extract a key from correlations obtained by measuring quantum systems. Device-independent quantum key distribution (DIQKD) is based on the observation that when these correlations violate a Bell inequality, a secure key can be extracted even if the users' devices are not fully characterised~\cite{pironio09,scarani12,arnonfriedman18,ekert91}. In a DIQKD security proof, it is merely assumed that the devices do not signal to the adversary or other components except when foreseen by the protocol~\cite{pironio09,scarani12,arnonfriedman18}. 
This differs from traditional QKD protocols~\cite{bennett84}, which are device-dependent in that they assume the devices are implementing operations within specified tolerances~\cite{scarani09}. Implementations of such protocols have been attacked by various methods~\cite{fung07,gerhardt11,jain15}, which exploit imperfections that cause the devices to operate outside the prescribed models. By working with fewer assumptions, DIQKD can achieve secure key distribution without detailed device characterisation, which would make the systems more reliable against such attacks. 

Unfortunately, there has been substantial difficulty in finding security proofs for DIQKD protocols with sufficient noise tolerance for physical implementation. One approach towards improving the tolerance is to investigate the information-reconciliation step. In QKD, the raw data of the users is not perfectly correlated, and they need to agree on a shared key using public communication. Existing DIQKD security proofs~\cite{pironio09,arnonfriedman18} have used one-way error-correction protocols in this step. However, for classical key reconciliation~\cite{maurer93,wolfthesis} and device-dependent QKD~\cite{gottesman03,chau02,rennerthesis,bae07,watanabe07,khatri17}, the noise tolerance can be improved by using two-way communication, a concept that has been referred to as \emph{advantage distillation}. 

It is natural to ask whether this concept could be extended to DIQKD. However, device-dependent security proofs for advantage distillation are often based on detailed state characterisations, given by measurements that are tomographically complete or nearly so~\cite{gottesman03,chau02,rennerthesis,bae07,watanabe07,khatri17}. This is generally not available in noisy DIQKD scenarios, where there can be many states and measurements compatible with the observed statistics. While recent works~\cite{kaur18,winczewski19} have found upper bounds on DIQKD key rates even with two-way communication, there do not appear to be any achievability results resolving the question of whether two-way communication provides an advantage in DIQKD.

In this work, we answer this question in the affirmative, by showing that advantage distillation yields better noise tolerances than one-way error correction in several scenarios. Our key observation is that even with the limited state characterisation available in DIQKD, it is still possible to identify and bound some important parameters that can be used in a security proof. We present our results in the form of several sufficient conditions for advantage distillation to be secure, together with a semidefinite programming (SDP) method to verify when these conditions hold. 

Our security proof is valid in the \emph{collective-attacks} regime, where one assumes all states and measurements are independent and identically distributed (IID) across the protocol rounds, but the adversary Eve can store quantum side-information and perform joint measurements on her collected states~\cite{scarani09}. Other attack models include \emph{individual attacks}, where Eve has no quantum memory, or the most general \emph{coherent attacks}, where the IID assumption is removed. Collective attacks can be stronger than individual attacks~\cite{gisin99,bae07}, but are often no weaker than coherent attacks~\cite{scarani09,arnonfriedman18}; we focus on collective attacks here. 

We focus on improving the asymptotic noise-tolerance thresholds, i.e.~the maximum noise at which key generation is still possible in the limit of many rounds. This is an important parameter when considering a proof-of-principle realisation of DIQKD.
\setcounter{footnote}{9}
Our approach also yields lower bounds on the asymptotic key rate~\footnote{See Supplemental Material, which includes Refs.~\cite{winter16,roga10,vazirani14,tomamichel17,yang14,fuchs95,gisin07,andersson17,scarani08,hubner92,briet09}}.

\emph{Conditions for security} --- 
Consider a DIQKD protocol between two parties Alice and Bob, where Alice has $\mathcal{X}$ possible measurements $A_0, A_1, ..., A_{\mathcal{X}-1}$, and similarly Bob has $\mathcal{Y}$ possible measurements $B_0, B_1, ..., B_{\mathcal{Y}-1}$, with $A_0, B_0$ taken to be binary-outcome measurements that generate a raw key. Eve holds a purification $E$ of Alice and Bob's states, and under the collective-attack assumption the states and measurements are IID, so we can focus on the single-round Alice-Bob-Eve state $\rho_{ABE}$. We assume that the devices do not eventually broadcast the final key through methods such as device-reuse attacks~\cite{barrett13} or covert channels~\cite{curty19}. This assumption could be supported by implementing measures such as those proposed in~\cite{barrett13,mckague14,lacerda19,curty19}.

Given the IID structure, parameter estimation can be performed to arbitrary accuracy, so we shall assume the outcome probabilities $\pr_{AB|XY}(ab|xy)$ for all measurement pairs $(A_x, B_y)$ are fully characterised in the protocol. (We will suppress subscripts for probability distributions when they are clear from context.) 
For convenience in the proofs, we assume a \emph{symmetrisation} step is implemented, in which Alice generates a uniform random bit $T$ in each round and sends it to Bob, with both parties flipping their measurement outcome if and only if $T=1$~\footnote{In this work, we take the symmetrisation step to be applied to all measurements, which is possible because we focus on scenarios where all measurements have binary outcomes. In principle, one could instead symmetrise the key-generating measurements only, in which case the measurements not used for key generation can have more outcomes.
}. The bit $T$ can be absorbed into Eve's side-information $E$. (This symmetrisation step can be omitted in practice; see~\cite{Note10} Sec.~\appsymm.) 
After this process, the measurements $A_0$ and $B_0$ have symmetrised outcomes, in the sense $\pr(01|00)=\pr(10|00)=\e/2$ and $\pr(00|00)=\pr(11|00)=(1-\e)/2$ for some $\e < 1/2$~\footnote{If $\e>1/2$, simply swap Bob's outcome labels.}. 
Henceforth, $\pr_{AB|XY}$ refers to the distribution after symmetrisation. 

We focus on the repetition-code protocol~\cite{maurer93,wolfthesis,rennerthesis,bae07} for advantage distillation, which is based on a block of $n$ rounds in which $A_0$ and $B_0$ were measured (we shall denote the output bitstrings as $\str{A}_0$ and $\str{B}_0$, and Eve's side-information across all the rounds as $\str{E}$). Alice privately generates a uniformly random bit $C$, and sends the message $\str{M} = \str{A}_0 \oplus (C,C,...,C)$ to Bob via a public authenticated channel. Bob replies with a bit $D$ that expresses whether to accept the block, with $D=1$ (accept) if and only if $\str{B}_0 \oplus \str{M} = (C',C',...,C')$ for some $C' \in \mathbb{Z}_2$. If the resulting systems satisfy
\begin{equation}
r \coloneqq H(C|\str{E}\str{M};D=1) - H(C|C';D=1) > 0,
\label{eq_keyrate}
\end{equation}
where $H$ is the von Neumann entropy, then repeating this procedure over many $n$-round blocks would allow a secret key to be distilled asymptotically from the bits $(C,C')$ in the accepted blocks~\cite{devetak05,arnonfriedman18}. Excluding parameter-estimation rounds, the key rate will be $r(\e^n + (1-\e)^n)/n$~\cite{rennerthesis}.

We derive~\cite{Note10} the following theorem (where $F(\rho,\sigma) = \norm{\sqrt{\rho} \sqrt{\sigma}}_1$ is the root-fidelity):
\begin{theorem}
\label{th_fidbound}
For a DIQKD protocol as described above, a sufficient condition for Eq.~\eqref{eq_keyrate} to hold for large $n$ is
\begin{align}
F(\rho_{E|00},\rho_{E|11})^2 > \frac{\e}{1-\e},
\label{eq_fidbound}
\end{align}
where $\rho_{E|a_0 b_0}$ is Eve's single-round state conditioned on $(A_0,B_0)$ being measured with outcome $(a_0,b_0)$.
\end{theorem}

The intuition behind the proof is that if Eve sees the message value $\str{M} = \str{m}$, then with high probability Alice and Bob's strings have the value $\str{A}_0 \str{B}_0 = \str{m} \str{m}$ or $\flip{\str{m}} \flip{\str{m}}$ (where $\flip{\str{m}} \defvar \str{m}\oplus\str{1}$). Hence Eve essentially has to distinguish between these two cases, which can be quantified via the fidelity $F(\rho_{\str{E}|\str{m} \str{m}},\rho_{\str{E}|\flip{\str{m}} \flip{\str{m}}}) = F(\rho_{E|00},\rho_{E|11})^n$.

Eq.~\eqref{eq_fidbound} is similar to the condition obtained in~\cite{bae07} for device-dependent QKD, but it is derived here without detailed state characterisation. However, it still remains to find bounds on $F(\rho_{E|00},\rho_{E|11})$ without device-dependent assumptions. We approach this task by combining the Fuchs--van de Graaf inequality~\cite{fuchs99} with the operational interpretation of trace distance:
\begin{equation}
\begin{aligned}
F(\rho_{E|00},\rho_{E|11}) \geq&\, 1-d(\rho_{E|00},\rho_{E|11}) \\
=&\, 2 (1-P_g(\rho_{E|00},\rho_{E|11})),
\end{aligned}
\label{eq_fvdg} 
\end{equation}
where $P_g(\rho_{E|00},\rho_{E|11})$ is Eve's maximum probability of guessing $C$ given the $E$ part of a c-q state $\sigma_{CE} = \sum_c (1/2) \ketbra{c}{c} \otimes \rho_{E|cc}$. In a DIQKD protocol as described above, $P_g(\rho_{E|00},\rho_{E|11})$ can be viewed as Eve's guessing probability for the outcome of $A_0 B_0$, conditioned on the outcome being $00$ or $11$. A DI method to bound such guessing probabilities based on the distribution $\pr_{AB|XY}$ was described in~\cite{thinh16}, using the family of SDPs known as the NPA hierarchy~\cite{navascues08}. We can hence apply this method to find whether Eq.~\eqref{eq_fidbound} holds for various distributions.

\begin{table*}
\caption{Noise thresholds for advantage distillation in various DIQKD scenarios. $\prt$ is the ideal probability distribution that the devices should implement in the absence of noise, and $\qt$ is the maximum depolarising noise such that we can show positive key rate is achievable using Theorem~\ref{th_fidbound} (for rows (i)--(iii)) or Corollary~\ref{co_22weakbound} (for rows (iv)--(vi)). Analogously, $\etat$ is the minimum efficiency which can be tolerated when we instead consider a limited-detection-efficiency model. Unless otherwise specified, the state used for $\prt$ is $\ket{\Phi^+} = (\ket{00} + \ket{11})/\sqrt{2}$.}
\def\arraystretch{1.5} 
\setlength\tabcolsep{.28cm}
\begin{center}
\begin{tabular}{C{5.3cm} C{8.5cm} C{.8cm} C{.9cm}}
\toprule
Description of $\prt$ & State and measurements for $\prt$ & $\qt$ & $\etat$ \\
\hline
\setcounter{tabrownumcounter}{0}
(\tabrownum) Achieves maximal CHSH value with the measurements $A_0,A_1,B_1,B_2$. & \makecell{$A_0=B_0=Z,\enspace A_1=X,$ \\ $B_1=(X+Z)/\sqrt{2},\enspace B_2=(X-Z)/\sqrt{2}$.} & 6.0\% & 93.7\% \\
\hline
(\tabrownum) Modification of a distribution exhibiting the Hardy paradox~\cite{hardy93,rabelo12} for improved robustness against limited detection efficiency. & $\ket{\psi} = \sqrt{\kappa}(\ket{01} + \ket{10}) + \sqrt{1-2\kappa} \ket{11}$ with $\kappa=(3-\sqrt{5})/2$; the $0$ outcomes correspond to projectors onto $\ket{a_0}=\ket{b_0}\propto\left(\sqrt{1+2\kappa}-\sqrt{1-2\kappa}\right)\ket{0} + 2\sqrt{\kappa}\ket{1}$, $\ket{a_1}=\ket{b_1}\approx 0.37972\ket{0} + 0.92510\ket{1}$, $\ket{a_2}=\ket{b_2}\approx 0.90821\ket{0} +  0.41851\ket{1}$. & 3.2\% & 92.0\% \\
\hline
(\tabrownum) Includes the Mayers-Yao self-test~\cite{mayers04} and the CHSH measurements. & \makecell{$A_0=B_0= Z,\enspace A_1=B_1= (X+Z)/\sqrt{2},$\\ $A_2=B_2= X,\enspace A_3=B_3= (X-Z)/\sqrt{2}$.} & 6.8\% & 92.7\% \\
\hline
(\tabrownum) Achieves maximal CHSH value with the measurements $A_0,A_1,B_0,B_1$. & \makecell{$A_0=Z,\enspace A_1=X,$\\ $B_0=(X+Z)/\sqrt{2},\enspace B_1=(X-Z)/\sqrt{2}$.} & 7.7\% & 91.7\% \\
\hline
(\tabrownum) Similar to (\scenCHSHbasic), but with measurements optimised for robustness against depolarising noise. & Measurements are in the $x$-$z$ plane at angles $\theta_{A_0} = 0.4187,\enspace \theta_{A_1} = 1.7900,\enspace \theta_{B_0} = 0.8636,\enspace \theta_{B_1} = 2.6340$. & 9.1\% & 90.0\% \\
\hline
(\tabrownum) Similar to (\scenCHSHbasic), but with states and measurements maximising CHSH violation for each value of detection efficiency $\eta$~\cite{eberhard93}. & $\ket{\psi} = \cos\Omega \ket{00} + \sin\Omega \ket{11}$ with $\Omega = 0.6224$; the 0 outcomes correspond to projectors onto states of the form $\cos(\theta/2)\ket{0} + \sin(\theta/2)\ket{1}$ with $\theta_{A_0} = -\theta_{B_0} = -0.35923,\enspace \theta_{A_1} = -\theta_{B_1} = 1.1538$. & 7.3\% & 89.1\% \\
\toprule
\end{tabular}
\end{center}
\def\arraystretch{1}
\label{tab_thresh}
\end{table*}

However, Eq.~\eqref{eq_fvdg} is generally not an optimal bound. We observe that if $\rho_{E|00}$ and $\rho_{E|11}$ were both assumed to be pure, then it could be replaced by a better relation,
\begin{equation}
F(\rho_{E|00},\rho_{E|11})^2 = 1-d(\rho_{E|00},\rho_{E|11})^2.
\label{eq_satfvdg}
\end{equation}
While it seems difficult to justify such an assumption in general, we show that for 2-input 2-output protocols, one can almost replace Eq.~\eqref{eq_fvdg} with Eq.~\eqref{eq_satfvdg} after taking a particular concave envelope~\cite{Note10}: 

\begin{theorem}
\label{th_22bound}
Consider a DIQKD protocol as described above, with $\mathcal{X}=\mathcal{Y}=2$ and all measurements having binary outcomes. Denoting the set of quantum distributions with $\pr(00|00)=\pr(11|00)$ as $\symmset$, let $f$ be a concave function on $\symmset$ such that for any $\gamma \in \symmset$, all states and measurements compatible with $\gamma$ satisfy $f(\gamma) \geq (1-\e)d(\rho_{E|00},\rho_{E|11})^2$. Then a sufficient condition for Eq.~\eqref{eq_keyrate} to hold for large $n$ is
\begin{equation}
{1-\frac{f(\pr_{AB|XY})}{1-\e}} > {\frac{\e}{1-\e}}.
\label{eq_22bound}
\end{equation}
\end{theorem}

Currently, we do not have a method for finding an optimal concave bound on $(1-\e)d(\rho_{E|00},\rho_{E|11})^2$. However, we find a condition that is more restrictive than Eq.~\eqref{eq_22bound} but more tractable to verify:

\begin{corollary}
\label{co_22weakbound}
Consider a DIQKD protocol as described above, with $\mathcal{X}=\mathcal{Y}=2$ and all measurements having binary outcomes. Then a sufficient condition for Eq.~\eqref{eq_keyrate} to hold for large $n$ is
\begin{equation}
{1-d(\rho_{E|00},\rho_{E|11})} > {\frac{\e}{1-\e}}.
\label{eq_22weakbound}
\end{equation}
\end{corollary}

As before, we can bound $d(\rho_{E|00},\rho_{E|11})$ by using the NPA hierarchy. Effectively, Corollary~\ref{co_22weakbound} improves over the combination of Theorem~\ref{th_fidbound} and Eq.~\eqref{eq_fvdg} by replacing $\left(1-d(\rho_{E|00},\rho_{E|11})\right)^2$ with $1-d(\rho_{E|00},\rho_{E|11})$. 

\emph{Noise thresholds} --- Using this method, we study the effects of two possible noise models for binary-outcome distributions. The first is depolarising noise parametrised by $\q \in [0,1/2]$:
\begin{equation}
\pr(ab|xy) = (1-2\q) \prt(ab|xy) + \q/2,
\label{eq_noise}
\end{equation}
where $\prt$ is some ideal target distribution~\footnote{Here $\pr(ab|xy)$ refers to the probabilities before symmetrisation, though if $\prt$ already has symmetrised outcomes then symmetrising $\pr(ab|xy)$ has no further effect.}. The second noise model is limited detection efficiency parametrised by $\eta \in [0,1]$, where all outcomes are subjected to independent Z-channels that flip $1$ to $0$ with probability $1-\eta$. This is a standard model for photonic setups where photon loss or non-detection occurs with probability $1-\eta$, with such events assigned to outcome $0$~\cite{pironio09}. ($\eta$ is an effective parameter describing all such losses. Given more detailed noise models~\cite{caprara15}, our method can be applied to the resulting distributions for more precise results.)

\begin{figure*}
\subfloat[Depolarising noise]{\label{fig_22CHSHa}
\includegraphics[width=0.45\textwidth]{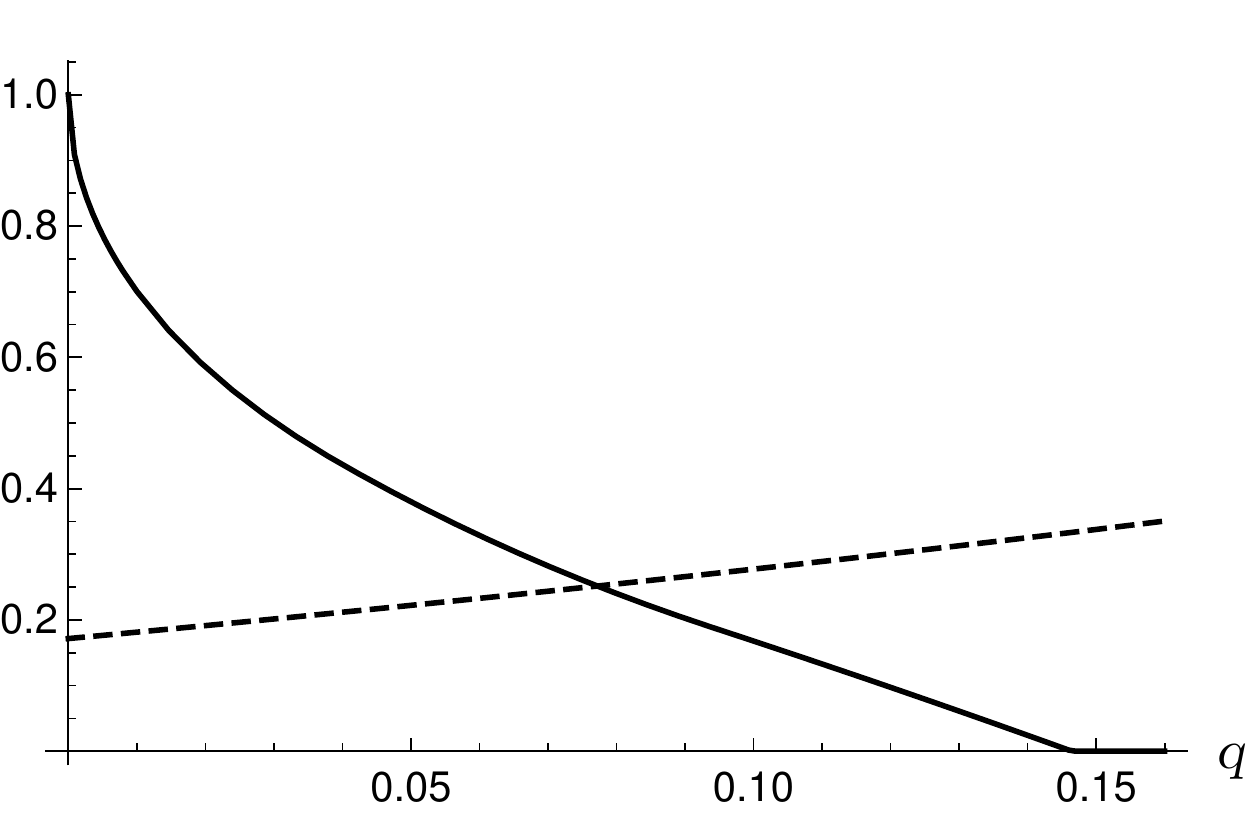}
} \hspace{.5cm}
\subfloat[Limited detection efficiency]{\label{fig_22CHSHb}
\includegraphics[width=0.45\textwidth]{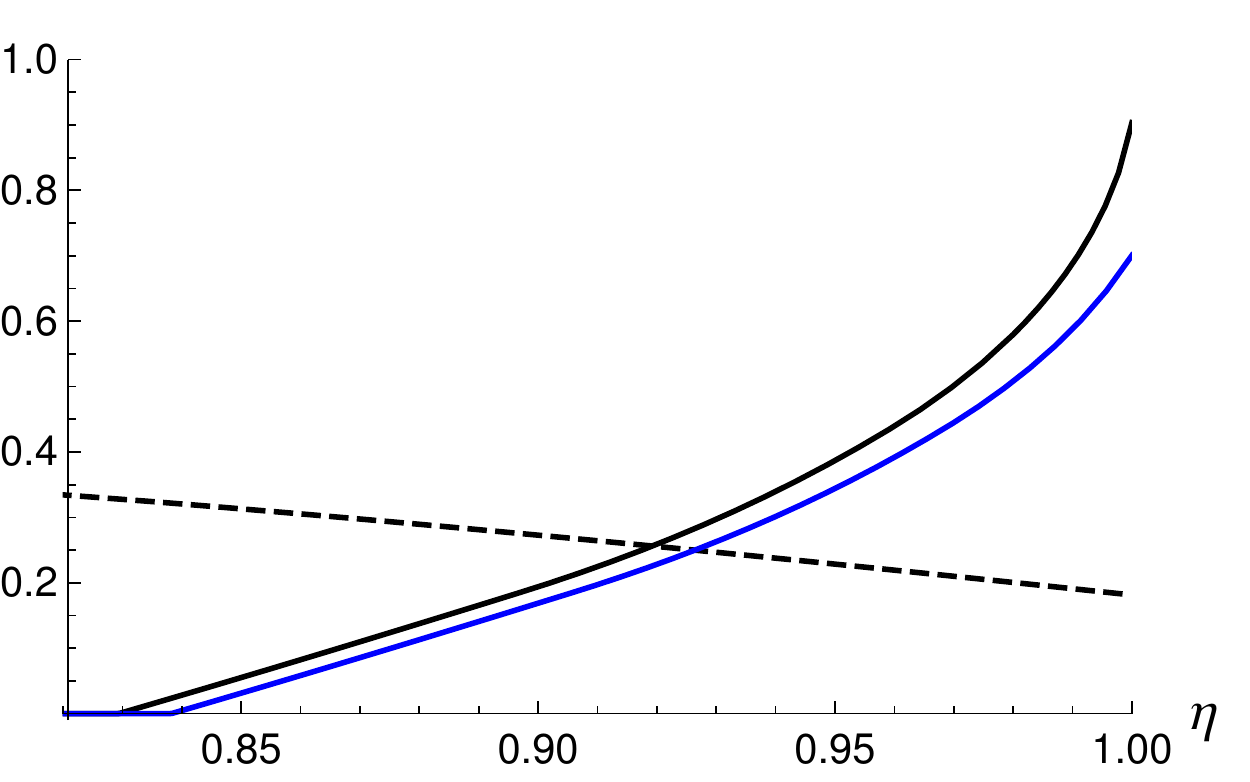}
}
\caption{Left- and right-hand sides of Eq.~\eqref{eq_22weakbound}, shown as solid and dashed curves respectively, for a DIQKD scenario where the target distribution attains maximum CHSH violation in the absence of noise. Plot~(a) shows the effect of depolarising noise $\q$, while in plot~(b) a small amount of depolarising noise is applied ($\q = 0.1\%$ for the black curve, $\q = 1\%$ for the blue curve) followed by a limited-detection-efficiency noise model with efficiency $\eta$. In plots~(a) and (b), the solid and dashed curves intersect at $\qt \approx 7.7\%$ and $\etat \approx 91.7\%\text{ (black), } 92.6\%\text{ (blue)}$ respectively, which yield the threshold values such that we can show positive key rate is achievable via Corollary~\ref{co_22weakbound}. The solid curves reach zero at the same noise values as where the CHSH violation becomes zero.
}
\label{fig_22CHSH}
\end{figure*}

In Table~\ref{tab_thresh}, we present a selection of our results (see~\cite{Note10} for the full list). 
Additionally, in Fig.~\ref{fig_22CHSH} we plot both sides of Eq.~\eqref{eq_22weakbound} for row~(\scenCHSHbasic) of the table. 
From the table, we see that appropriate choices of $\prt$ can tolerate depolarising noise of $\qt \approx 9.1\%$ or detection efficiencies of $\etat \approx 89.1\%$. This indeed outperforms the DIQKD protocol in~\cite{pironio09} based on one-way error correction, which can tolerate $\qt \approx 7.1\%$ or $\etat \approx 92.4\%$ (or $\etat \approx 90.7\%$ for a modified version where the state and measurements $A_0, A_1, B_1, B_2$ are optimised to maximise the CHSH value for each value of $\eta$~\cite{eberhard93}, and $B_0$ is then separately optimised to be maximally correlated to $A_0$). 

We observe that the DIQKD protocol in~\cite{pironio09} uses essentially the same $\prt$ as row~(\scenoneway) in Table~\ref{tab_thresh}. This is not a 2-input 2-output scenario, and so the noise thresholds we can prove for that specific setup are somewhat worse. However, row~(\scenCHSHbasic) is in fact the same scenario with one measurement \emph{omitted}, making it a 2-input 2-output scenario, thus we could use Corollary~\ref{co_22weakbound} to show that advantage distillation in this scenario can surpass the thresholds in~\cite{pironio09}. Hence we have shown that for the scenario in~\cite{pironio09}, advantage distillation achieves a higher noise tolerance even while ignoring one measurement. This is particularly surprising since the key-generating measurements in row~(\scenCHSHbasic) are not perfectly correlated. In fact, if the proof in~\cite{pironio09} were applied to this scenario~\footnote{By replacing the error-correction term $H(A_0|B_0) = \binh(\q)$ with $H(A_0|B_0) = \binh(\e)$.}, it would only tolerate noise up to $\qt \approx 3.1\%$.
If we instead allow optimisation of the states and measurements for noise robustness, then the relevant rows are (\scenCHSHq) and (\scenCHSHeta), where the noise thresholds we find for advantage distillation also outperform one-way error correction.

In Table~\ref{tab_thresh}, the noise thresholds for scenarios with more than 2 inputs are generally worse, because for such scenarios we cannot apply Corollary~\ref{co_22weakbound}. The best results we have for such cases are listed in rows (\scenHardyopt) and (\scenMY). It would be of interest to find a way to overcome this issue, perhaps by finding more direct bounds on $F(\rho_{E|00},\rho_{E|11})$, or further study of when the analysis can be reduced to states satisfying Eq.~\eqref{eq_satfvdg}.
We observe that pure states are not the only states satisfying the equation --- for instance, if $\rho_{E|00}$ and $\rho_{E|11}$ are qubit states, the equality holds if and only if they have the same eigenvalues (see~\cite{Note10} Sec.~\appfvdg). 

\emph{Conclusion and outlook} --- 
In summary, we have found that by using advantage distillation, the noise thresholds for DIQKD with one-way error correction can be surpassed. Specifically, advantage distillation is secure against collective attacks up to depolarising-noise values of $q \approx 9.1\%$ or detection efficiencies of $\eta \approx 89.1\%$, which exceeds the best-known noise thresholds of $q \approx 7.1\%$ and $\eta \approx 90.7\%$ respectively for DIQKD with one-way error correction. 

Currently, we require large block sizes $n$ to certify positive key rates. However, small block sizes are sufficient for reasonable asymptotic key rates in the device-dependent case~\cite{rennerthesis}. Tighter bounds on $F(\rho_{E|00},\rho_{E|11})$ should give similar results in DIQKD, hence this would be an important next step. Alternatively, one could analyse the finite-key security~\cite{tomamichel12,arnonfriedman18,murta19}. Since our approach yields explicit bounds~\cite{Note10} on the entropies in Eq.~\eqref{eq_keyrate}, it could in principle be extended to a finite-size security proof against collective attacks by using the quantum asymptotic equipartition property~\cite{tomamichel09}, following the approach in~\cite{murta19}. However, this approach is likely to require a large number of rounds to achieve positive key rates, which would pose a challenge for practical implementation.

Another significant goal would be extending our results to non-IID scenarios. We conjecture that allowing coherent attacks will not change the asymptotic key rates, as was the case for various device-dependent QKD and DIQKD protocols~\cite{scarani09,arnonfriedman18}. To support this, we observe that if the measurements have an IID tensor-product structure, then the analysis of any permutation-symmetric protocol can be asymptotically reduced to the IID case using de Finetti theorems~\cite{rennerthesis}, assuming the system dimensions are bounded. Hence any attack that can be modelled by simply using non-IID states (with IID measurements) cannot yield an asymptotic advantage over collective attacks (see~\cite{Note10} Sec.~\appattack). To find a security proof for non-IID measurements, the entropy accumulation theorem~\cite{dupuis16,arnonfriedman18} or a new type of de Finetti theorem may be required.

Finally, an open question in information theory is the existence of \emph{bound information}, referring to correlations which require secret bits to be produced but from which no secret key can be extracted~\cite{gisin00,khatri17}. There is a simple analogue to this in the context of DIQKD, namely whether there exist correlations which violate Bell inequalities but cannot be distilled into a secret key in a DI setting. Our results have a gap between the noise thresholds at which we can no longer prove the protocol's security and the thresholds at which the Bell violation becomes zero (see also~\cite{Note10} Sec.~\appattack, where we outline a potential attack for $\q \gtrsim 12.8\%$ if the users only measure $\e$ and the CHSH value). It would be of interest to find whether this gap can be closed. \\

\begin{acknowledgments}
We thank Jean-Daniel Bancal, Srijita Kundu, Joseph Renes, Valerio Scarani, Le Phuc Thinh and Marco Tomamichel for helpful discussions. E.\ Y.-Z.\ Tan and R.\ Renner were funded by the Swiss National Science Foundation via the National Center for Competence in Research for Quantum Science and Technology (QSIT), and by the Air Force Office of Scientific Research (AFOSR) via grant FA9550-19-1-0202. C.\ C.-W.\ Lim acknowledges funding support from the National Research Foundation of Singapore:  NRF Fellowship grant (NRFF11-2019-0001) and NRF Quantum Engineering Programme grant (QEP-P2) and the Centre for Quantum Technologies, and the Asian Office of Aerospace Research and Development. Computations were performed with the NPAHierarchy function in QETLAB~\cite{qetlab}, using the CVX package~\cite{cvxpackage,cvxbook} with solver SDPT3.
\end{acknowledgments}

\bibliography{biblio_repcodeDI} 



\section*{Supplementary material}

# Supplemental Material

## A. PROOF DETAILS

For completeness, we repeat the protocol description: consider the outputs of a block of $n$ rounds in which $A_0$ and $B_0$ were measured (we shall denote the output bitstrings as $\boldsymbol{A}_0$ and $\boldsymbol{B}_0$, and Eve's side-information across all the rounds as $\boldsymbol{E}$). Alice privately generates a uniformly random bit $C$, and sends the message $\boldsymbol{M} = \boldsymbol{A}_0 \oplus (C, C, ..., C)$ to Bob via a public authenticated channel. Bob replies with a single bit $D$ that expresses whether to accept the block, with $D = 1$ (accept) if and only if $\boldsymbol{B}_0 \oplus \boldsymbol{M} = (C', C', ..., C')$ for some $C' \in \mathbb{Z}_2$. If the resulting systems satisfy

$$H(C|\boldsymbol{E}\boldsymbol{M}; D = 1) - H(C|C'; D = 1) > 0, \tag{1}$$

then repeating this procedure over many $n$-round blocks would allow a secret key to be distilled asymptotically from the bitstrings $\boldsymbol{C}, \boldsymbol{C'}$ in the accepted blocks [S1, S2]. We assume a symmetrisation step is implemented, and denote the resulting distribution as $\Pr_{AB|XY}$. The statements and proofs of the theorems are as follows:

**Theorem 1.** *For a DIQKD protocol as described above, a sufficient condition for Eq. (1) to hold for large $n$ is*

$$F(\rho_{E|00}, \rho_{E|11})^2 > \frac{\epsilon}{1-\epsilon}, \tag{2}$$

*where $\rho_{E|a_0 b_0}$ is Eve's single-round state conditioned on $(A_0, B_0)$ being measured with outcome $(a_0, b_0)$.*

*Proof.* All states denoted in this proof are normalised. To bound $H(C|\boldsymbol{E}\boldsymbol{M}; D = 1)$, we first observe that since $H(X|YZ) = \sum_z \Pr_Z(z) H(X|Y; Z = z)$ for classical $Z$, it suffices to bound $H(C|\boldsymbol{E}; \boldsymbol{M} = \boldsymbol{m} \wedge D = 1)$ for arbitrary messages $\boldsymbol{m}$. Starting from the initial $\boldsymbol{A}_0 \boldsymbol{B}_0 \boldsymbol{E}$ state

$$\rho_{\boldsymbol{A}_0 \boldsymbol{B}_0 \boldsymbol{E}} = \sum_{\boldsymbol{a}_0, \boldsymbol{b}_0} \Pr_{\boldsymbol{A}_0 \boldsymbol{B}_0}(\boldsymbol{a}_0, \boldsymbol{b}_0) |\boldsymbol{a}_0, \boldsymbol{b}_0\rangle\langle \boldsymbol{a}_0, \boldsymbol{b}_0| \otimes \rho_{\boldsymbol{E}|\boldsymbol{a}_0 \boldsymbol{b}_0}, \tag{S1}$$

a straightforward calculation shows that conditioned on the block being accepted and $\boldsymbol{M} = \boldsymbol{m}$, the $C\boldsymbol{E}$ state takes the form $\rho_{C\boldsymbol{E}|\boldsymbol{M}=\boldsymbol{m} \wedge D=1} = \sum_c (1/2) |c\rangle\langle c| \otimes \omega_c$ with

$$\omega_0 = \frac{\Pr_{\boldsymbol{A}_0 \boldsymbol{B}_0}(\boldsymbol{m}, \boldsymbol{m}) \rho_{\boldsymbol{E}|\boldsymbol{m}\boldsymbol{m}} + \Pr_{\boldsymbol{A}_0 \boldsymbol{B}_0}(\boldsymbol{m}, \overline{\boldsymbol{m}}) \rho_{\boldsymbol{E}|\boldsymbol{m}\overline{\boldsymbol{m}}}}{\Pr_{\boldsymbol{A}_0 \boldsymbol{B}_0}(\boldsymbol{m}, \boldsymbol{m}) + \Pr_{\boldsymbol{A}_0 \boldsymbol{B}_0}(\boldsymbol{m}, \overline{\boldsymbol{m}})}, \qquad \omega_1 = \frac{\Pr_{\boldsymbol{A}_0 \boldsymbol{B}_0}(\overline{\boldsymbol{m}}, \overline{\boldsymbol{m}}) \rho_{\boldsymbol{E}|\overline{\boldsymbol{m}}\overline{\boldsymbol{m}}} + \Pr_{\boldsymbol{A}_0 \boldsymbol{B}_0}(\overline{\boldsymbol{m}}, \boldsymbol{m}) \rho_{\boldsymbol{E}|\overline{\boldsymbol{m}}\boldsymbol{m}}}{\Pr_{\boldsymbol{A}_0 \boldsymbol{B}_0}(\overline{\boldsymbol{m}}, \overline{\boldsymbol{m}}) + \Pr_{\boldsymbol{A}_0 \boldsymbol{B}_0}(\overline{\boldsymbol{m}}, \boldsymbol{m})}, \tag{S2}$$

where $\overline{\boldsymbol{m}} = \boldsymbol{m} \oplus \boldsymbol{1}$. We now consider a state $\tilde{\rho}_{C\boldsymbol{E}}$ defined as follows:

$$\tilde{\rho}_{C\boldsymbol{E}} = (1/2)(|0\rangle\langle 0| \otimes \rho_{\boldsymbol{E}|\boldsymbol{m}\boldsymbol{m}} + |1\rangle\langle 1| \otimes \rho_{\boldsymbol{E}|\overline{\boldsymbol{m}}\overline{\boldsymbol{m}}}). \tag{S3}$$

With symmetrised IID outcomes, we have $\Pr_{\boldsymbol{A}_0 \boldsymbol{B}_0}(\boldsymbol{m}, \boldsymbol{m}) = \Pr_{\boldsymbol{A}_0 \boldsymbol{B}_0}(\overline{\boldsymbol{m}}, \overline{\boldsymbol{m}}) = (1-\epsilon)^n/2^n$ and $\Pr_{\boldsymbol{A}_0 \boldsymbol{B}_0}(\boldsymbol{m}, \overline{\boldsymbol{m}}) = \Pr_{\boldsymbol{A}_0 \boldsymbol{B}_0}(\boldsymbol{m}, \overline{\boldsymbol{m}}) = \epsilon^n/2^n$, so

$$d(\tilde{\rho}_{C\boldsymbol{E}}, \rho_{C\boldsymbol{E}|\boldsymbol{M}=\boldsymbol{m}\wedge D=1}) \leq \delta_n, \text{ where } \delta_n = \frac{\epsilon^n}{\epsilon^n + (1-\epsilon)^n}. \tag{S4}$$

Applying a continuity bound for conditional von Neumann entropy [S3] then yields

$$H(C|\boldsymbol{E}; \boldsymbol{M} = \boldsymbol{m} \wedge D = 1) \geq H(C|\boldsymbol{E})_{\tilde{\rho}} - \delta_n - (1 + \delta_n) h_2\left(\frac{\delta_n}{1 + \delta_n}\right), \tag{S5}$$

where $h_2$ is the binary entropy function. The $H(C|\boldsymbol{E})_{\tilde{\rho}}$ term is bounded by [S4][S5]

$$H(C|\boldsymbol{E})_{\tilde{\rho}} \geq 1 - h_2\left(\frac{1 - F(\rho_{\boldsymbol{E}|\boldsymbol{m}\boldsymbol{m}}, \rho_{\boldsymbol{E}|\overline{\boldsymbol{m}}\overline{\boldsymbol{m}}})}{2}\right) = 1 - h_2\left(\frac{1 - F(\rho_{E|00}, \rho_{E|11})^n}{2}\right), \tag{S6}$$

using the IID assumption. As for $H(C|C'; D = 1)$, it can be seen that $\Pr(C \neq C' | D = 1) = \delta_n$ for IID outcomes, so

$$H(C|C'; D = 1) = h_2(\delta_n). \tag{S7}$$

Combining these results, we conclude

$$\frac{H(C|\boldsymbol{E}\boldsymbol{M}; D = 1)}{H(C|C'; D = 1)} \geq \left(1 - h_2\left(\frac{1 - F(\rho_{E|00}, \rho_{E|11})^n}{2}\right) - \delta_n - (1 + \delta_n) h_2\left(\frac{\delta_n}{1 + \delta_n}\right)\right) h_2(\delta_n)^{-1}. \tag{S8}$$

It remains to find the behaviour of this expression in the large-$n$ limit. To do so, we first note that letting $\phi(\delta) = h_2(\delta/(1+\delta))$, we have $\phi(\delta), h_2(\delta) \to 0$ and $\phi'(\delta), h'_2(\delta) \to \infty$ as $\delta \to 0^+$, so

$$\lim_{\delta \to 0^+} \frac{\delta}{h_2(\delta)} = \lim_{\delta \to 0^+} \frac{1}{h'_2(\delta)} = 0, \tag{S9}$$

$$\lim_{\delta \to 0^+} \frac{\phi(\delta)}{h_2(\delta)} = \lim_{\delta \to 0^+} \frac{\phi''(\delta)}{h''_2(\delta)} = \lim_{\delta \to 0^+} \frac{(1+\delta - 2\delta \ln \delta)/(\delta(1+\delta)^3)}{1/(\delta(1-\delta))} = 1. \tag{S10}$$

Since $\delta_n \to 0^+$ as $n \to \infty$, the terms arising from the continuity bound hence have a finite limit,

$$\lim_{n \to \infty} \left( -\delta_n - (1+\delta_n)h_2\left(\frac{\delta_n}{1+\delta_n}\right) \right) h_2(\delta_n)^{-1} = -1. \tag{S11}$$

Let $\beta = \epsilon/(1-\epsilon) \in [0, 1)$. Then $\delta_n \leq \beta^n$, and for sufficiently large $n$ we have $\beta^n < 1/2$, so $h_2(\delta_n) \leq h_2(\beta^n) \leq 2\beta^n \log(1/\beta^n)$. Hence for any $\alpha > \beta$,

$$\lim_{n \to \infty} \frac{\alpha^n}{h_2(\delta_n)} \geq \lim_{n \to \infty} \frac{\alpha^n}{2n\beta^n \log(1/\beta)} = \infty. \tag{S12}$$

The right-hand side of Eq. (S8) can be lower-bounded by using $h_2((1-p)/2) \leq 1 - p^2/\ln 4$ (this inequality follows from the Taylor expansion of $h_2$) to get $1 - h_2\left((1 - F(\rho_{E|00}, \rho_{E|11})^n)/2\right) \geq F(\rho_{E|00}, \rho_{E|11})^{2n}/\ln 4$. Combining this with the choice $\alpha = F(\rho_{E|00}, \rho_{E|11})^2$ in Eq. (S12), we conclude that when Eq. (2) holds, the right-hand side of Eq. (S8) limits to $\infty$ as $n \to \infty$. Therefore, Eq. (1) will hold for sufficiently large $n$. $\square$

Given specific values or bounds for $F(\rho_{E|00}, \rho_{E|11}), \epsilon, n$, one can substitute them into Eq. (S8) to get an explicit bound on the asymptotic key rate. Note that $\rho_{E|00}$ and $\rho_{E|11}$ refer to the states after the symmetrisation step is implemented. The theorem can in fact be slightly broadened to encompass protocols without a symmetrisation step, as long as $\Pr_{AB|XY}$ has symmetrised outcomes, since the latter property is all that is required in this proof. However, since in any case the symmetrisation step can be omitted in practice via the analysis in Sec. C, this does not seem to be a significant generalisation.

This security proof has a key difference in structure compared to a proof technique often used for QKD with one-way error correction, for instance in [S2, S6, S7]. Specifically, in this proof we must consider the "security" of both key-generating measurements $(A_0, B_0)$, while the proofs for QKD with one-way error correction often only need to bound the security of $A_0$ (in terms of the smoothed min-entropy of the bitstring it produces). Broadly speaking, the approach in those proofs works because when Alice sends the error-correction string to Bob, the information leakage to Eve is bounded simply by the length of the string, which is fixed beforehand based on the expected *honest* behaviour of the devices. For the repetition-code protocol, however, a large number of bits are publicly communicated, and hence bounding the information leakage via the number of bits yields too crude a bound. The need to explicitly consider the security of $B_0$ in this protocol can also be seen by considering an extreme example where Eve always knows the outcome of $B_0$, possibly at the cost of it being poorly correlated to $A_0$. In that case, regardless of how secure the output of $A_0$ is, Eve will always know the value of $C'$, making it impossible to distill key from the $(C, C')$ pairs. Hence any security proof for this protocol must involve some kind of security argument regarding $B_0$, even if only indirectly via measuring its correlations with $A_0$.

In the main text, we used the Fuchs-van de Graaf inequality to bound the fidelity $F(\rho_{E|00}, \rho_{E|11})$, but this is likely not optimal, given the fairly large gap between the upper and lower bounds of the inequality. It might potentially be possible to directly bound $F(\rho_{E|00}, \rho_{E|11})$ by relating it to higher moments of the NPA hierarchy, in the vein of one approach for robust self-testing [S8]. Alternatively, one could perhaps use the fact that for any pair of states $(\rho, \sigma)$, there exists a measurement such that the resulting outcome distributions $(P, Q)$ satisfy $F(P, Q) = F(\rho, \sigma)$ [S9]. This effectively allows a reduction to classical side-information, in which case the minimum value of $F(P, Q)$ given the observed $\Pr_{AB|XY}$ can be phrased as an optimisation problem analogous to [S10].

**Theorem 2.** *Consider a DIQKD protocol as described above, with $\mathcal{X} = \mathcal{Y} = 2$ and all measurements having binary outcomes. Denoting the set of quantum distributions with $\Pr(00|00) = \Pr(11|00)$ as $\mathcal{S}$, let $f$ be a concave function on $\mathcal{S}$ such that for any $\gamma \in \mathcal{S}$, all states and measurements compatible with $\gamma$ satisfy $f(\gamma) \geq (1-\epsilon)d(\rho_{E|00}, \rho_{E|11})^2$. Then a sufficient condition for Eq. (1) to hold for large $n$ is*

$$1 - \frac{f(\Pr_{AB|XY})}{1-\epsilon} > \frac{\epsilon}{1-\epsilon}. \tag{5}$$

*Proof.* We use the fact that in a 2-input 2-output DI protocol, Jordan's lemma can be used to argue [S11, S12] that without loss of generality, we can assume Eve's strategy in each round consists of generating a random variable $\Lambda$ and storing it in a classical register, then implementing some corresponding *qubit strategy*, which is a strategy such that $\rho_{ABE|\lambda}$ is a $2 \times 2 \times 4$ pure state and all of Alice and Bob's measurements are rank-1 projective measurements. For a qubit strategy, Eve's state conditioned on the joint outcome of both measurements is pure. Hence if Alice and Bob measure $(A_x, B_y)$ and store their results in classical registers $(\bar{A}_x, \bar{B}_y)$, we can take the resulting single-round state $\rho_{\bar{A}_x \bar{B}_y E\Lambda}$ after symmetrisation to be in the form [S13]

$$\rho_{\bar{A}_x \bar{B}_y E\Lambda} = \sum_{a,b} \sum_{\lambda} \Pr(ab|xy\lambda) \Pr(\lambda) |a,b\rangle\langle a,b| \otimes \rho_{E|\lambda ab} \otimes |\lambda\rangle\langle\lambda|, \tag{S13}$$

where the probabilities $\Pr(ab|xy\lambda)$ satisfy $\sum_\lambda \Pr(ab|xy\lambda)\Pr(\lambda) = \Pr(ab|xy)$, and for each $\lambda$ we have $\Pr(00|00\lambda) = \Pr(11|00\lambda) = (1-\epsilon_\lambda)/2$ for some $\epsilon_\lambda \in [0,1]$, due to the symmetrisation step. Using the reduction to qubit strategies, we can also take all $\rho_{E|\lambda ab}$ to be pure states (see Sec. B). When Alice and Bob both perform the key-generating measurements and get the same outcome (which we shall denote here as $\gamma \in \{0,1\}$), Eve's conditional states are

$$\rho_{E\Lambda|\gamma\gamma} = \sum_\lambda \widetilde{\Pr}(\lambda) \rho_{E|\lambda\gamma\gamma} \otimes |\lambda\rangle\langle\lambda|, \text{ where } \widetilde{\Pr}(\lambda) = \frac{\Pr(\gamma\gamma|00\lambda)}{\Pr(\gamma\gamma|00)} \Pr(\lambda) = \frac{1-\epsilon_\lambda}{1-\epsilon} \Pr(\lambda). \tag{S14}$$

Note that $\widetilde{\Pr}$ is itself a valid probability distribution over $\Lambda$, and independent of $\gamma$.

We now apply the same arguments as in the proof of Theorem 1 up until Eq. (S5), though in this case Eve's side-information takes the form $\boldsymbol{E\Lambda}$, and so we consider the state $\tilde{\rho}_{C\boldsymbol{E\Lambda}} = (1/2)(|0\rangle\langle 0| \otimes \rho_{\boldsymbol{E\Lambda}|\boldsymbol{mm}} + |1\rangle\langle 1| \otimes \rho_{\boldsymbol{E\Lambda}|\overline{\boldsymbol{mm}}})$. Using the above, one can show that

$$\tilde{\rho}_{C\boldsymbol{E\Lambda}} = \sum_{\boldsymbol{\lambda}} \widetilde{\Pr}(\boldsymbol{\lambda}) \frac{1}{2} \left( |0\rangle\langle 0| \otimes \rho_{\boldsymbol{E}|\boldsymbol{\lambda mm}} + |1\rangle\langle 1| \otimes \rho_{\boldsymbol{E}|\boldsymbol{\lambda}\overline{\boldsymbol{mm}}} \right) \otimes |\boldsymbol{\lambda}\rangle\langle\boldsymbol{\lambda}|, \tag{S15}$$

where $\widetilde{\Pr}(\boldsymbol{\lambda}) = \prod_j \widetilde{\Pr}(\lambda_j)$ and $\rho_{\boldsymbol{E}|\boldsymbol{\lambda\gamma\gamma}} = \bigotimes_j \rho_{E|\lambda_j \gamma_j \gamma_j}$. Hence we have $H(C|\boldsymbol{E\Lambda})_{\tilde{\rho}} = \sum_{\boldsymbol{\lambda}} \widetilde{\Pr}(\boldsymbol{\lambda}) H(C|\boldsymbol{E}; \boldsymbol{\Lambda} = \boldsymbol{\lambda})_{\tilde{\rho}}$. Applying the bound from [S4] together with $h_2((1-p)/2) \leq 1 - p^2/\ln 4$ then yields

$$H(C|\boldsymbol{E\Lambda})_{\tilde{\rho}} \geq \sum_{\boldsymbol{\lambda}} \widetilde{\Pr}(\boldsymbol{\lambda}) \left( 1 - h_2\left( \frac{1 - F(\rho_{\boldsymbol{E}|\boldsymbol{\lambda mm}}, \rho_{\boldsymbol{E}|\boldsymbol{\lambda}\overline{\boldsymbol{mm}}})}{2} \right) \right) \geq \frac{1}{\ln 4} \sum_{\boldsymbol{\lambda}} \widetilde{\Pr}(\boldsymbol{\lambda}) F(\rho_{\boldsymbol{E}|\boldsymbol{\lambda mm}}, \rho_{\boldsymbol{E}|\boldsymbol{\lambda}\overline{\boldsymbol{mm}}})^2. \tag{S16}$$

Using the IID structure,

$$\sum_{\boldsymbol{\lambda}} \widetilde{\Pr}(\boldsymbol{\lambda}) F(\rho_{\boldsymbol{E}|\boldsymbol{\lambda mm}}, \rho_{\boldsymbol{E}|\boldsymbol{\lambda}\overline{\boldsymbol{mm}}})^2 = \sum_{\boldsymbol{\lambda}} \prod_{j=1}^n \widetilde{\Pr}(\lambda_j) F(\rho_{E|\lambda_j 00}, \rho_{E|\lambda_j 11})^2 = \left( \sum_\lambda \widetilde{\Pr}(\lambda) F(\rho_{E|\lambda 00}, \rho_{E|\lambda 11})^2 \right)^n. \tag{S17}$$

Since the states $\rho_{E|\lambda 00}, \rho_{E|\lambda 11}$ are pure, they satisfy $F(\rho_{E|00}, \rho_{E|11})^2 = 1 - d(\rho_{E|00}, \rho_{E|11})^2$, and so

$$\sum_\lambda \widetilde{\Pr}(\lambda) F(\rho_{E|\lambda 00}, \rho_{E|\lambda 11})^2 = 1 - \sum_\lambda \frac{1-\epsilon_\lambda}{1-\epsilon} \Pr(\lambda) d(\rho_{E|\lambda 00}, \rho_{E|\lambda 11})^2 \geq 1 - \frac{1}{1-\epsilon} \sum_\lambda \Pr(\lambda) f(\Pr_{AB|XY\lambda}). \tag{S18}$$

Finally, this is lower-bounded by $1 - f(\Pr_{AB|XY})/(1-\epsilon)$ since $f$ is concave, so

$$\frac{H(C|\boldsymbol{E\Lambda M}; D=1)}{H(C|C'; D=1)} \geq \left( \frac{1}{\ln 4} \left( 1 - \frac{f(\Pr_{AB|XY})}{1-\epsilon} \right)^n - \delta_n - (1+\delta_n) h_2\left( \frac{\delta_n}{1+\delta_n} \right) \right) h_2(\delta_n)^{-1}. \tag{S19}$$

The behaviour of this expression in the large-$n$ limit is given by the same analysis as the last part of the proof of Theorem 1, choosing $\alpha = 1 - f(\Pr_{AB|XY})/(1-\epsilon)$ in this case. We conclude that when Eq. (5) holds, the right-hand side of Eq. (S19) limits to $\infty$ as $n \to \infty$. Therefore, Eq. (1) will hold for sufficiently large $n$. □

Similar to Theorem 1, values can be substituted into Eq. (S19) to obtain explicit bounds on the key rate. Unlike Theorem 1, the proof of this theorem does require an explicit symmetrisation step (see Sec. B), though the analysis in Sec. C allows one to avoid implementing the symmetrisation step in practice. If $f$ is the *optimal* concave upper bound on $(1-\epsilon) d(\rho_{E|00}, \rho_{E|11})^2$, in the sense that there always exists a mixture of qubit strategies such that $\sum_\lambda \Pr(\lambda)(1-\epsilon_\lambda) d(\rho_{E|\lambda 00}, \rho_{E|\lambda 11})^2 = f(\Pr_{AB|XY})$, then the above analysis is essentially tight for large $n$. This is because in Eq. (S16), the first inequality [S4] is in fact saturated because $\rho_{\boldsymbol{E}|\boldsymbol{\lambda mm}}, \rho_{\boldsymbol{E}|\boldsymbol{\lambda}\overline{\boldsymbol{mm}}}$ are pure, and the second inequality is approximately saturated at large $n$ because $h_2((1-p)/2) = 1 - p^2/\ln 4 - O(p^4)$.

**Corollary 1.** *Consider a DIQKD protocol as described above, with $\mathcal{X} = \mathcal{Y} = 2$ and all measurements having binary outcomes. Then a sufficient condition for Eq. (1) to hold for large $n$ is*

$$1 - d(\rho_{E|00}, \rho_{E|11}) > \frac{\epsilon}{1-\epsilon}. \tag{6}$$

*Proof.* Let $\tilde{f}$ be the optimal upper bound on $(1-\epsilon)d(\rho_{E|00}, \rho_{E|11})$, defined on the set of quantum distributions such that $\Pr(00|00) = \Pr(11|00)$. Since $d(\rho_{E|00}, \rho_{E|11}) \leq 1$, we have [S14]

$$\tilde{f}(\Pr_{AB|XY}) \geq (1-\epsilon)d(\rho_{E|00}, \rho_{E|11}) \geq (1-\epsilon)d(\rho_{E|00}, \rho_{E|11})^2. \tag{S20}$$

Also, $\tilde{f}$ must be concave, for essentially the same reason that optimal guessing-probability bounds must be concave, though modified to account for the "postselection" on the outcomes being 00 or 11. More specifically, denote the set of quantum distributions with $\Pr(00|00) = \Pr(11|00)$ as $\mathcal{S}$. Consider any probability distribution $\Pr(\lambda)$ and any family of distributions $\Pr_{AB|XY\lambda} \in \mathcal{S}$ indexed by $\lambda$, and take an arbitrary $\delta > 0$. For each $\lambda$, there exists a strategy for Eve that achieves the distribution $\Pr_{AB|XY\lambda}$ and has $(1-\epsilon_\lambda)d(\rho_{E|\lambda 00}, \rho_{E|\lambda 11}) \geq \tilde{f}(\Pr_{AB|XY\lambda}) - \delta$, because $\tilde{f}$ is an optimal bound (i.e. there exist strategies arbitrarily close to saturating the bound). If Eve generates and stores a classical random variable $\Lambda$ according to the distribution $\Pr(\lambda)$, then implements the corresponding strategy, the resulting states take the same form as in Eq. (S13),(S14). This is a strategy that achieves probabilities $\Pr(ab|xy) = \sum_\lambda \Pr(\lambda)\Pr(ab|xy\lambda)$, and therefore

$$\begin{aligned}
\tilde{f}(\Pr_{AB|XY}) &\geq (1-\epsilon)d(\rho_{E|00}, \rho_{E|11}) \\
&= (1-\epsilon)d\left(\sum_\lambda \widetilde{\Pr}(\lambda)\rho_{E|\lambda 00} \otimes |\lambda\rangle\langle\lambda|, \sum_\lambda \widetilde{\Pr}(\lambda)\rho_{E|\lambda 11} \otimes |\lambda\rangle\langle\lambda|\right) \\
&= (1-\epsilon)\sum_\lambda \widetilde{\Pr}(\lambda) d\left(\rho_{E|\lambda 00}, \rho_{E|\lambda 11}\right) \\
&= \sum_\lambda \Pr(\lambda)(1-\epsilon_\lambda) d\left(\rho_{E|\lambda 00}, \rho_{E|\lambda 11}\right) \\
&\geq \left(\sum_\lambda \Pr(\lambda)\tilde{f}(\Pr_{AB|XY\lambda})\right) - \delta.
\end{aligned}$$

Since $\delta$ was arbitrary, we conclude that $\tilde{f}(\Pr_{AB|XY}) \geq \sum_\lambda \Pr(\lambda)\tilde{f}(\Pr_{AB|XY\lambda})$, i.e. $\tilde{f}$ is concave on $\mathcal{S}$. Hence choosing $f = \tilde{f}$ satisfies the conditions of Theorem 2. Since $\tilde{f}$ is an upper bound on $(1-\epsilon)d(\rho_{E|00}, \rho_{E|11})$, we conclude that when Eq. (6) holds, we have

$$1 - \frac{\tilde{f}(\Pr_{AB|XY})}{1-\epsilon} \geq 1 - d(\rho_{E|00}, \rho_{E|11}) > \frac{\epsilon}{1-\epsilon}, \tag{S21}$$

and the claim follows by Theorem 2. □

## B. EFFECT OF SYMMETRISATION

The argument essentially follows the reduction to Bell-diagonal states in [S11], with slight modifications to account for the fact that we are bounding fidelity rather than entropy. Denote the single-round state before symmetrisation as $\sigma$. As mentioned in the proof of Theorem 2, we use Jordan's lemma to reduce the analysis to a classical mixture of qubit strategies, and since each party has only two measurements, we can choose coordinates so that all measurements lie in the $x$-$z$ plane. When Alice and Bob measure $A_x B_y$ and store their results in registers $\bar{A}_x, \bar{B}_y$, the resulting single-round state $\sigma_{\bar{A}_x \bar{B}_y E\Lambda}$ is

$$\sigma_{\bar{A}_x \bar{B}_y E\Lambda} = \sum_{a,b} \sum_\lambda \Pr_i(ab|xy\lambda) \Pr(\lambda) |a,b\rangle\langle a,b| \otimes \sigma_{E|\lambda ab} \otimes |\lambda\rangle\langle\lambda|, \tag{S22}$$

where all $\sigma_{E|\lambda ab}$ are pure states, and $\Pr_i$ refers to the initial probability distribution before symmetrisation. After the symmetrisation is carried out via a publicly communicated bit $T$, we have the state

$$\rho_{\bar{A}_x \bar{B}_y ET\Lambda} = \sum_{a,b} \sum_\lambda \Pr(ab|xy\lambda) \Pr(\lambda) |a,b\rangle\langle a,b| \otimes \rho_{ET|\lambda ab} \otimes |\lambda\rangle\langle\lambda|,$$

$$\text{where } \Pr(ab|xy\lambda) = \sum_t \frac{1}{2} \Pr_i(a \oplus t, b \oplus t | xy\lambda), \quad \rho_{ET|\lambda ab} = \sum_t \frac{\Pr_i(a \oplus t, b \oplus t | xy\lambda)}{2\Pr(ab|xy\lambda)} \sigma_{E|\lambda, a \oplus t, b \oplus t} \otimes |t\rangle\langle t|.$$

(S23)

The probabilities $\Pr(ab|xy\lambda)$ satisfy $\sum_\lambda \Pr(ab|xy\lambda)\Pr(\lambda) = \Pr(ab|xy)$, and for each value of $\lambda$, we have $\Pr(00|00\lambda) = \Pr(11|00\lambda)$. In the proof of Theorem 2, we should be considering the state $\rho$ with Eve having access to $E$ and $T$, and eventually need to bound $F(\rho_{ET|\lambda 00}, \rho_{ET|\lambda 11})$, but we would face the problem that the states $\rho_{ET|\lambda ab}$ are not pure. However, we shall now argue that there exists a state $\rho'$ which produces the same probabilities $\Pr(ab|xy\lambda)$ and achieves $F(\rho'_{ET|\lambda 00}, \rho'_{ET|\lambda 11}) = F(\rho_{ET|\lambda 00}, \rho_{ET|\lambda 11})$ with *pure* conditional states $\rho'_{ET|\lambda ab}$, essentially by replacing the classical register $T$ with an appropriate purification. Therefore, we can consider the state $\rho'$ instead of the state $\rho$ when bounding $F(\rho_{ET|\lambda 00}, \rho_{ET|\lambda 11})$.

Denoting the pure pre-measurement states $\sigma_{ABE|\lambda}$ as $|\psi\rangle\langle\psi|_{ABE|\lambda}$, consider the state

$$\rho'_{ABET\Lambda} = \sum_\lambda \Pr(\lambda) |\psi'\rangle\langle\psi'|_{ABET|\lambda} \otimes |\lambda\rangle\langle\lambda|,$$

$$\text{where } |\psi'\rangle_{ABET|\lambda} = \frac{1}{\sqrt{2}} \left( |\psi\rangle_{ABE|\lambda} \otimes |0\rangle + (Y \otimes Y \otimes \text{id}) |\psi\rangle_{ABE|\lambda} \otimes |1\rangle \right),$$

(S24)

and let Alice and Bob's devices perform the same measurements as they did on $\sigma$. Denote the post-measurement states after measuring $A_x B_y$ on $|\psi\rangle_{ABE|\lambda}$ and getting outcome $ab$ as $|\psi\rangle_{E|\lambda ab}$. Since $Y \otimes Y$ acting on Alice and Bob's systems has the effect of flipping the outcomes of all Pauli measurements in the $x$-$z$ plane, the post-measurement state can be found to be

$$\rho'_{\bar{A}_x \bar{B}_y ET\Lambda} = \sum_{a,b} \sum_\lambda \Pr(ab|xy\lambda) \Pr(\lambda) |a,b\rangle\langle a,b| \otimes |\psi'\rangle\langle\psi'|_{ET|\lambda ab} \otimes |\lambda\rangle\langle\lambda|,$$

$$\text{where } \Pr(ab|xy\lambda) = \sum_t \frac{1}{2} \Pr_i(a \oplus t, b \oplus t | xy\lambda), \quad |\psi'\rangle_{ET|\lambda ab} = \sum_t \sqrt{\frac{\Pr_i(a \oplus t, b \oplus t | xy\lambda)}{2\Pr(ab|xy\lambda)}} |\psi\rangle_{E|\lambda, a \oplus t, b \oplus t} \otimes |t\rangle.$$

(S25)

This state achieves the same probabilities $\Pr(ab|xy\lambda)$ as $\rho$ in Eq. (S23). In addition, we can verify that the pure states $\rho'_{ET|\lambda ab} = |\psi'\rangle\langle\psi'|_{ET|\lambda ab}$ satisfy $F(\rho'_{ET|\lambda 00}, \rho'_{ET|\lambda 11}) = F(\rho_{ET|\lambda 00}, \rho_{ET|\lambda 11})$. Therefore, we can consider this $\rho'$ in place of $\rho$.

## C. OMITTING THE SYMMETRISATION STEP

To show that the symmetrisation step is not required in practice, we consider two scenarios. The first scenario describes what happens when a symmetrisation step is applied before carrying out the repetition-code protocol, while the second scenario describes a case where the repetition-code protocol is carried out without the symmetrisation step, followed by a few additional operations. By comparing these two scenarios, we can argue that the repetition-code protocol without the symmetrisation step is still secure.

Let Scenario 1 be defined as follows. Alice and Bob initially have bitstrings $\boldsymbol{A}_0, \boldsymbol{B}_0$ of length $n$, and Eve has side-information $\boldsymbol{E}$. Alice generates a uniformly random bitstring $\boldsymbol{T}$ of length $n$, and sends it to Bob via a public channel. They use $\boldsymbol{T}$ to compute new bitstrings $\tilde{\boldsymbol{A}}_0 = \boldsymbol{A}_0 \oplus \boldsymbol{T}$, $\tilde{\boldsymbol{B}}_0 = \boldsymbol{B}_0 \oplus \boldsymbol{T}$, and then perform the repetition-code protocol on the bitstrings $\tilde{\boldsymbol{A}}_0, \tilde{\boldsymbol{B}}_0$ (which now have symmetrised outcomes). In this case, the message Alice sends is $\tilde{\boldsymbol{M}} = \tilde{\boldsymbol{A}}_0 \oplus (C, C, ..., C)$, and Bob computes $\tilde{\boldsymbol{B}}_0 \oplus \tilde{\boldsymbol{M}}$. For ease of argument in the next steps, we shall suppose that Alice also privately computes $\boldsymbol{M} = \boldsymbol{A}_0 \oplus (C, C, ..., C)$. At the end of this process, the state capturing all the relevant information takes the form $\rho_{\boldsymbol{A}_0 \boldsymbol{B}_0 \tilde{\boldsymbol{A}}_0 \tilde{\boldsymbol{B}}_0 \boldsymbol{ETM}\tilde{\boldsymbol{M}} CC'D}$, with Eve having access to $\boldsymbol{ET}\tilde{\boldsymbol{M}}D$. (To ensure $C'$ is always well-defined, we can assign $C'$ to take some null value $\perp$ when $\tilde{\boldsymbol{B}}_0 \oplus \tilde{\boldsymbol{M}}$ is not a string of $n$ bits with the same value.)

Now let Scenario 2 be a hypothetical scenario where first, Alice and Bob perform the repetition-code protocol directly on the bitstrings $\boldsymbol{A}_0, \boldsymbol{B}_0$, without symmetrisation. After this is done, Alice generates a uniformly random bitstring $\boldsymbol{T}$ and sends it to Bob via a public channel, then Alice and Bob use $\boldsymbol{T}$ to privately compute $\tilde{\boldsymbol{A}}_0, \tilde{\boldsymbol{B}}_0$ and $\tilde{\boldsymbol{M}}$ the same way as in Scenario 1. At the end of this process, the state capturing all the relevant information takes the

form $\rho_{\boldsymbol{A}_0\boldsymbol{B}_0\tilde{\boldsymbol{A}}_0\tilde{\boldsymbol{B}}_0\boldsymbol{ETM}\tilde{\boldsymbol{M}}CC'D}$, and the key observation is that this state is exactly the same in both Scenario 1 and Scenario 2. (The main point to check is whether $C'$ and $D$ have the same behaviour in both scenarios, which can be verified by noting that in fact $\tilde{\boldsymbol{B}}_0 \oplus \tilde{\boldsymbol{M}} = \boldsymbol{B}_0 \oplus \boldsymbol{M}$.) However, in this case Eve instead has access to $\boldsymbol{ETM}D$.

We now note that the three variables $\boldsymbol{T}, \boldsymbol{M}, \tilde{\boldsymbol{M}}$ have the property that given the values of any two of them, the value of the third is fixed. Therefore, we can say that $H(C|\boldsymbol{ETM}; D=1) = H(C|\boldsymbol{ETM}\tilde{\boldsymbol{M}}; D=1) = H(C|\boldsymbol{ET}\tilde{\boldsymbol{M}}; D=1)$. (It does not matter whether we are referring to Scenario 1 or Scenario 2 here, since the states describing both scenarios are identical.) Finally, we can see in Scenario 2 that $\boldsymbol{T}$ is clearly independent of $C\boldsymbol{E}D$, since it was only generated after the repetition-code protocol was performed, independently of all systems involved in that process. Therefore, we must have $H(C|\boldsymbol{ETM}; D=1) = H(C|\boldsymbol{EM}; D=1)$ as well.

To find the values of $H(C|\boldsymbol{EM}; D=1)$ and $H(C|C'; D=1)$ for the protocol without symmetrisation, we simply need to consider their values on the final state in Scenario 2, since all the steps carried out after the repetition-code protocol is completed in Scenario 2 do not change these entropies. As we have argued above, $H(C|\boldsymbol{EM}; D=1)$ in Scenario 2 equals $H(C|\boldsymbol{ET}\tilde{\boldsymbol{M}}; D=1)$ in Scenario 1, and so the first term is the same in the protocols with and without symmetrisation. In addition, since the states described in the two scenarios are identical, the values of $H(C|C'; D=1)$ are also equal.

Overall, this means that the value of $H(C|\boldsymbol{EM}; D=1) - H(C|C'; D=1)$ for the protocol without symmetrisation is the same as the value of $H(C|\boldsymbol{ET}\tilde{\boldsymbol{M}}; D=1) - H(C|C'; D=1)$ for the protocol with symmetrisation. Therefore, bounding the latter already proves the security of the protocol without symmetrisation, and so our theorems also apply even without the symmetrisation step. However, note that when applying our theorems in that context, the bounds on $F(\rho_{E|00}, \rho_{E|11})$ or the functions $f$ and $g$ must not be computed based on the unsymmetrised $\Pr_{AB|XY}$ obtained directly from parameter estimation, but rather based on the hypothetical distribution that would be obtained after symmetrisation (which can be easily computed from the unsymmetrised $\Pr_{AB|XY}$).

## D. FULL LIST OF NOISE THRESHOLDS

TABLE S1. Noise thresholds for advantage distillation in various DIQKD scenarios, given by Theorem 1 (for rows (i)–(vi)) or Corollary 1 (for rows (vii)–(x)). Unless otherwise specified, the state used for $\Pr_{\text{target}}$ is $|\Phi^+\rangle = (|00\rangle + |11\rangle)/\sqrt{2}$.

| Description of $\Pr_{\text{target}}$ | State and measurements for $\Pr_{\text{target}}$ | $q_t$ | $\eta_t$ |
|---|---|---|---|
| (i) Achieves maximal CHSH value with the measurements $A_0, A_1, B_1, B_2$. | $A_0 = B_0 = Z,\ A_1 = X,$ $B_1 = (X+Z)/\sqrt{2},\ B_2 = (X-Z)/\sqrt{2}.$ | 6.0% | 93.7% |
| (ii) Exhibits the Hardy paradox "maximally" [S15, S16] on $A_1, A_2, B_1, B_2$, with $A_0, B_0$ in the Schmidt basis. | $|\psi\rangle = \sqrt{\kappa}(|01\rangle + |10\rangle) + \sqrt{1-2\kappa}\,|11\rangle$ with $\kappa = (3-\sqrt{5})/2$; the 0 outcomes correspond to projectors onto $|a_0\rangle = |b_0\rangle \propto \left(\sqrt{1+2\kappa} - \sqrt{1-2\kappa}\right)|0\rangle + 2\sqrt{\kappa}\,|1\rangle,$ $|a_1\rangle = |b_1\rangle = |1\rangle,\ |a_2\rangle = |b_2\rangle \propto \sqrt{\kappa}\,|0\rangle + \sqrt{1-2\kappa}\,|1\rangle.$ | 3.0% | 92.6% |
| (iii) Similar to (ii), but with measurements optimised for robustness against limited detector efficiency. | $|\psi\rangle = \sqrt{\kappa}(|01\rangle + |10\rangle) + \sqrt{1-2\kappa}\,|11\rangle$ with $\kappa = (3-\sqrt{5})/2$; the 0 outcomes correspond to projectors onto $|a_0\rangle = |b_0\rangle \propto \left(\sqrt{1+2\kappa} - \sqrt{1-2\kappa}\right)|0\rangle + 2\sqrt{\kappa}\,|1\rangle,$ $|a_1\rangle = |b_1\rangle \approx 0.37972\,|0\rangle + 0.92510\,|1\rangle,$ $|a_2\rangle = |b_2\rangle \approx 0.90821\,|0\rangle + 0.41851\,|1\rangle.$ | 3.2% | 92.0% |
| (iv) Includes the Mayers-Yao self-test [S17] and the CHSH measurements. | $A_0 = B_0 = Z,\ A_1 = B_1 = (X+Z)/\sqrt{2},$ $A_2 = B_2 = X,\ A_3 = B_3 = (X-Z)/\sqrt{2}.$ | 6.8% | 92.7% |
| (v) Maximally violates the elegant Bell inequality [S18, S19] with the measurements $A_1$–$A_3$, $B_0$–$B_3$. | $A_0 = B_0 = -(X+Y+Z)/\sqrt{3},$ $A_1 = Z,\ A_2 = Y,\ A_3 = X,\ B_1 = (X+Y-Z)/\sqrt{3},$ $B_2 = (X-Y+Z)/\sqrt{3},\ B_3 = (-X+Y+Z)/\sqrt{3}.$ | 6.1% | 93.6% |
| (vi) Maximally violates the elegant Bell inequality [S18, S19] with the measurements $A_0$–$A_2$, $B_1$–$B_4$. | $A_0 = B_0 = Z,\ A_1 = Y,\ A_2 = X,$ $B_1 = -(X+Y+Z)/\sqrt{3},\ B_2 = (X+Y-Z)/\sqrt{3},$ $B_3 = (X-Y+Z)/\sqrt{3},\ B_4 = (-X+Y+Z)/\sqrt{3}$ | 4.3% | 95.5% |
| (vii) Achieves maximal CHSH value with the measurements $A_0, A_1, B_0, B_1$. | $A_0 = Z,\ A_1 = X,$ $B_0 = (X+Z)/\sqrt{2},\ B_1 = (X-Z)/\sqrt{2}.$ | 7.7% | 91.7% |
| (viii) Similar to (vii), but with measurements optimised for robustness against depolarising noise. | Measurements are in the $x$-$z$ plane at angles $\theta_{A_0} = 0.4187,\ \theta_{A_1} = 1.7900,\ \theta_{B_0} = 0.8636,\ \theta_{B_1} = 2.6340.$ | 9.1% | 90.0% |
| (ix) Similar to (viii), but with the constraint $A_0 = B_0$. | Measurements are in the $x$-$z$ plane at angles $\theta_{A_0} = \theta_{B_0} = 0,\ \theta_{A_1} = \pi/3,\ \theta_{B_1} = 2\pi/3.$ | 7.1% | 92.3% |
| (x) Similar to (vii), but with states and measurements maximising CHSH violation for each value of detector efficiency $\eta$ [S20]. | $|\psi\rangle = \cos\Omega\,|00\rangle + \sin\Omega\,|11\rangle$ with $\Omega = 0.6224$; the 0 outcomes correspond to projectors onto states of the form $\cos(\theta/2)\,|0\rangle + \sin(\theta/2)\,|1\rangle$ with $\theta_{A_0} = -\theta_{B_0} = -0.35923,\ \theta_{A_1} = -\theta_{B_1} = 1.1538.$ | 7.3% | 89.1% |

The $\Pr_{\text{target}}$ distributions described in rows (ii),(iii),(x) do not have symmetrised outcomes, and hence their noise tolerances in the limited-detector-efficiency model are affected by the outcome labelling choice (the labellings chosen here appeared to be the optimal choices for those measurement angles). For these rows and all limited-detector-efficiency models, the distribution after applying noise does not have symmetrised outcomes, so a symmetrisation step was carried out after applying noise. Rows (i)–(iii) were computed at level 4 of the NPA hierarchy [S21][S22], rows (iv)–(vi) at level 3, and rows (vii)–(x) at level 5 (though in most cases, we found little difference from the values at level 2). Fig. 1 in the main text was computed at NPA level 2. Optimisations for noise robustness were carried out with fairly crude numerical methods, so the true optimal values may not have been found. We have not yet performed a very extensive search of possible choices for $\Pr_{\text{target}}$, so there is still room for exploration of other target distributions.

## E. DISCUSSION OF ATTACKS

### I. Beyond collective attacks

The repetition-code protocol may appear to have a weakness in that if Eve perfectly knows the outcomes $(A_0, B_0)$ for any single round in the block, then she also knows the values of $(C, C')$. However, we note that parameter estimation in the protocol should take place on a random subset of the rounds, so Eve cannot know precisely which rounds will be used for key generation (in any case, if she knew this in advance, she could simply choose to attack only the rounds where parameter estimation is not performed). Therefore, Eve cannot specifically target one round in each block. It is true that if Eve's strategy involves measuring a fraction $O(1/n)$ of the rounds to perfectly learn one outcome pair in each block with high probability, the effect on the estimated parameters is very small if a large block size $n$ is used. However, in principle this can still be detected asymptotically by using a large sample size (note that the block size $n$ is independent of the total number of rounds performed in the protocol, because it should be fixed before the protocol, based on the expected noise level).

More generally, when developing a security proof against non-IID attacks, one can typically include a step in the protocol where a random permutation is applied to the rounds, making the protocol permutation-symmetric. (This is purely to simplify the proofs, as the permutation step does not need to be performed in practice, via the arguments in [S23].) Parameter estimation by sampling a random subset is equivalent to sampling a fixed subset after the permutation step. For permutation-symmetric protocols with IID tensor-product measurements, quantum de Finetti theorems can be used to reduce the analysis to IID states, albeit with a dependence on the system dimensions. In the DI case, this implies that if we assume the system dimensions are finite and the measurements have a tensor-product structure, then the analysis of non-IID states can at least be reduced to IID states *asymptotically*, though an explicit bound on the system dimensions would be needed to obtain finite-size results with this approach. In particular, it would seem that the blockwise attack strategy described above could be modelled using non-IID states and IID tensor-product measurements, in which case it cannot give an asymptotic advantage over collective attacks as long as the protocol can be made permutation-symmetric.

### II. The CHSH scenario

Comparing to row (i) to row (vii) in Table S1, it would intuitively seem that using the additional perfectly correlated measurement in row (i) for key generation can only be beneficial. However, we observed a surprising behaviour in the scenario of row (i), namely that the bound on $d(\rho_{E|00}, \rho_{E|11})$ given by the SDP method becomes trivial even when there is still a CHSH violation from $A_0, A_1, B_1, B_2$. This contrasts with the behaviour for row (vii), where it appears that the bound on $d(\rho_{E|00}, \rho_{E|11})$ is nontrivial whenever the CHSH violation is nonzero (see Fig. 1 in the main text). A possible explanation is that since $B_0$ is not critical for Bell violation in row (i), it is less constrained and so Eve can have more information about its outcome. To support this notion, we now outline a potential attack strategy against the scenario in row (i) subject to depolarising noise, if the users only measure $\epsilon$ and the CHSH value.

For depolarising noise $q$, we have $\epsilon = q$ and $S = 2\sqrt{2}(1 - 2q)$ in this scenario, where $S$ denotes the CHSH value of $A_0, A_1, B_1, B_2$ when the outcomes are labelled as $\pm 1$. We shall show that for $q \geq 1/(5 + 2\sqrt{2})$, there exists an adversarial distribution $\Pr_{\mathrm{ad}}(ab|xy)$ that attains these values of $\epsilon$ and $S$, but allows Eve to perfectly distinguish the outcome 00 from the outcome 11 on the measurement pair $A_0B_0$. Consider a scenario where Eve generates a classical random variable $\Lambda \in \{0, 1, 2, 3\}$, and implements one of the following strategies based on the value of $\Lambda$:

$\Lambda = 0$: The devices implement the ideal singlet state and CHSH measurements for $A_0, A_1, B_1, B_2$, and $B_0$ is in the basis perfectly correlated to $A_0$. This gives $S = 2\sqrt{2}$ and $\epsilon = 0$.

$\Lambda = 1$: The devices implement the ideal singlet state and CHSH measurements for $A_0, A_1, B_1, B_2$, and $B_0$ is in the basis perfectly anticorrelated to $A_0$. This gives $S = 2\sqrt{2}$ and $\epsilon = 1$.

$\Lambda = 2$: The devices implement a classical (completely insecure) distribution attaining $S = 2$ and $\epsilon = 0$.

$\Lambda = 3$: The devices implement a classical (completely insecure) distribution attaining $S = 2$ and $\epsilon = 1$.

For $\Lambda \in \{0, 1\}$, Eve's state is independent of the outcome of $A_0B_0$, while for $\Lambda \in \{2, 3\}$, Eve perfectly knows the outcome of $A_0B_0$. The resulting distribution $\Pr_{\mathrm{ad}}(ab|xy)$ observed by Alice and Bob attains values of $S$ and $\epsilon$ that are related to the distribution of $\Lambda$ by $S = 2\sqrt{2}(\Pr_\Lambda(0) + \Pr_\Lambda(1)) + 2(\Pr_\Lambda(2) + \Pr_\Lambda(3))$, $\epsilon = \Pr_\Lambda(1) + \Pr_\Lambda(3)$. We now note that if $\Pr_\Lambda(0) = 0$, then whenever the outcome of $A_0B_0$ is 00 or 11, it must have been the case that $\Lambda \in \{2, 3\}$ (since the outcome of $A_0B_0$ is always 01 or 10 when $\Lambda = 1$), in which case the devices have implemented a completely

insecure distribution and so Eve knows the outcome perfectly. Finally, we observe that it is indeed possible to have $\Pr_\Lambda(0) = 0$ with $\epsilon = q$ and $S = 2\sqrt{2}(1-2q)$, by choosing

$$\Pr_\Lambda(1) = 1 - 2(2+\sqrt{2})q, \quad \Pr_\Lambda(2) = 1-q, \quad \Pr_\Lambda(3) = (5+2\sqrt{2})q - 1. \tag{S26}$$

All these values are nonnegative when $q \in [1/(5+2\sqrt{2}), 1/(2(2+\sqrt{2}))] \approx [12.8\%, 14.6\%]$, and hence form a valid probability distribution. Therefore, Eve can distinguish the 00 and 11 outcomes perfectly for $q \gtrsim 12.8\%$ (the value $q \approx 14.6\%$ is simply the noise value at which $S = 2$, so for higher noise values there is already no CHSH violation and no certification of security).

However, this attack focuses only on producing appropriate values of $\epsilon$ and $S$. It might already be possible to rule it out by considering the full distribution $\Pr_{AB|XY}$ instead. In addition, perfectly distinguishing the 00 and 11 outcomes is not precisely equivalent to a complete attack on the repetition-code protocol, since it is essentially ignoring what happens in the case where $\boldsymbol{A}_0 \neq \boldsymbol{B}_0$ in the accepted block, which occurs with small but nonzero probability $\delta_n$. Intuitively, this should not affect the efficacy of the attack since $\delta_n$ is small; also, $C \neq C'$ for such a block so Bob's guess $C'$ will be wrong in any case. We leave further analysis of this attack for future work. (Note that this attack cannot be straightforwardly applied to the 2-input 2-output scenarios in Table S1, because for those protocols the CHSH value includes $\langle A_0 B_0 \rangle$ directly, and it is impossible to have $S = 2\sqrt{2}$ and $\epsilon = 0$ simultaneously in that case.)

### F. SATURATING THE FUCHS-VAN DE GRAAF INEQUALITY

We find that for qubit states $\rho$ and $\sigma$, we have $F(\rho,\sigma)^2 + d(\rho,\sigma)^2 = 1$ if and only if $\rho$ and $\sigma$ have the same eigenvalues. To show this, we use the fact that letting the Bloch vectors of the qubit states be $\vec{u}$ and $\vec{v}$, we have the following formulas for root-fidelity [S24] and trace distance:

$$F(\rho,\sigma)^2 = \left(1 + \vec{u}\cdot\vec{v} + \sqrt{1-u^2}\sqrt{1-v^2}\right)/2, \quad d(\rho,\sigma)^2 = |\vec{u}-\vec{v}|^2/4 = (u^2+v^2-2\vec{u}\cdot\vec{v})/4, \tag{S27}$$

where $u = |\vec{u}|, v = |\vec{v}|$. Therefore,

$$F(\rho,\sigma)^2 + d(\rho,\sigma)^2 = \left(2 + 2\sqrt{1-u^2}\sqrt{1-v^2} + v^2 + u^2\right)/4, \tag{S28}$$

and solving for $F(\rho,\sigma)^2 + d(\rho,\sigma)^2 = 1$ yields the unique solution $u^2 = v^2$. Hence we conclude that for qubit states, we have $F(\rho,\sigma)^2 + d(\rho,\sigma)^2 = 1$ if and only if $\rho$ and $\sigma$ have Bloch vectors of the same length, which is equivalent to $\rho$ and $\sigma$ having the same eigenvalues.

This property does not seem to extend to non-qubit states: for instance, the family of qutrit states $\rho = p\left|0\right\rangle\!\left\langle 0\right| + (1-p)\left|1\right\rangle\!\left\langle 1\right|, \sigma = p\left|0\right\rangle\!\left\langle 0\right| + (1-p)\left|2\right\rangle\!\left\langle 2\right|$ with $p \in [0,1]$ instead satisfies $F(\rho,\sigma) = p$ and $d(\rho,\sigma) = 1-p$, which saturates the opposite bound in the Fuchs-van de Graaf inequality.

---



\end{document}